# Strong enhancement of photoresponsivity with shrinking the electrodes spacing in few layer GaSe photodetectors

Yufei Cao[1], Kaiming Cai[1], Pingan Hu[2], Lixia Zhao[3], Tengfei Yan[1], Xinhui Zhang[1], Xiaoguang Wu[1], Kaiyou Wang[1*] and Houzhi Zheng[1]

A critical challenge for the integration of the optoelectronics is that photodetectors have relatively poor sensitivities at the nanometer scale. It is generally believed that a large electrodes spacing in photodetectors is required to absorb sufficient light to maintain high photoresponsivity and reduce the dark current. However, this will limit the optoelectronic integration density. Through spatially resolved photocurrent investigation, we find that the photocurrent in metal-semiconductor-metal (MSM) photodetectors based on layered GaSe is mainly generated from the photoexcited carriers close to the metal-GaSe interface and the photocurrent active region is always close to the Schottky barrier with higher electrical potential. The photoresponsivity monotonically increases with shrinking the spacing distance before the direct tunneling happen, which was significantly enhanced up to 5,000 AW$^{-1}$ for the bottom

[1.] SKLSM, Institute of Semiconductors, Chinese Academy of Science, P. O. Box 912, Beijing 100083, P. R. China, [2.] KLMM, Harbin Institute of Technology, No. 2 YiKuang Street, Harbin, 150080, P.R. China, [3.] Semiconductor Lighting Technology Research and Development Center, Institute of Semiconductors, Chinese Academy of Sciences, Beijing, 100083, P. R. China. *E-mail: kywang@semi.ac.cn

contacted device at bias voltage 8 V and wavelength of 410 nm. It is more than 1,700-fold improvement over the previously reported results. Besides the systematically experimental investigation of the dependence of the photoresponsivity on the spacing distance for both the bottom and top contacted MSM photodetectors, a theoretical model has also been developed to well explain the photoresponsivity for these two types of device configurations. Our findings realize shrinking the spacing distance and improving the performance of 2D semiconductor based MSM photodetectors simultaneously, which could pave the way for future high density integration of 2D semiconductor optoelectronics with high performances.

The planar metal-semiconductor-metal (MSM) photodetectors based on layered materials have been studied extensively in recent years[1-5]. This is because the planar MSM photodetector has many advantages, such as compatibility with current semiconducting technology, very low dark current and high operation speed, which are attractive for many optoelectronic applications[6-8]. However, relative small photoresponsivity was firstly observed in these two dimensional (2D) layered materials based photodetectors including graphene (less than 0.1 $AW^{-1}$) and $MoS_2$ (no more than 7.5 $mAW^{-1}$)[4,9-11], which is because of the weak optical absorption or very small carrier mobility in these layered materials. In order to improve the photoresponsivity, graphene based photodetector has focused on enhancement of the absorption of light in graphene, for example by exploiting thermoelectric effects[12,13], microcavities[14,15] or multilayer tunneling structure[16] to improve its photoresponsivity up to 1,000 $AW^{-1}$. By improving the device mobility, the $MoS_2$ based photodetectors have reached a maximum external photoresponsivity of 880 AW$^{-1}$ (ref. 17). However, so far relative large device sizes were used in those investigations, impeding the high integration density applications. Since the importance of the device size and the photoresponsivity, we systematically investigated the relation between the photoresponsivity and the device size for both the top and bottom contacted MSM photodetectors based on layered GaSe. Combining the photocurrent measurements under global and spatially resolved illuminations[18], a model has been developed for understanding the underlying physics of photoresponsivity in our MSM photodetectors. Our work suggests that MSM photodetectors based on high photoresponse layered materials can be used for future high density optoelectronic applications.

The layered hexagonal GaSe was chosen to be the optical active material in this work because of its high photoresponsivity (2.8 AW$^{-1}$) and quantum efficiency (1,367%), which was demonstrated recently[19]. The GaSe crystals are composed of vertically stacked Se-Ga-Ga-Se sheets weakly bound by van der Waals interactions. Usually it is a p-type semiconductor (Supplementary Fig. S1) with an indirect bandgap of ~2.11 eV at the center of the Brillioun zone, which is only 25 meV above the conduction band minimum[19,20]. These two minima can both be populated by the photoexcited carriers, and then radiative recombinations from states associated with the direct and indirect gaps simultaneously occur. Thus, it causes the GaSe to be a promising material for optoelectronic applications. The GaSe crystal based optoelectronic devices can not only be used as photodetectors[19,21], THz source generator[22], but also for the nonlinear optical applications due to its large nonlinear optical coefficient (54 pmV$^{-1}$)[23].

**Results**

**Sample preparation and characterization.** Since the metal-semiconductor contact regime could play a very important role in the MSM photodetectors[24-26], the devices with two different types of design were fabricated based on mechanical exfoliated few layer GaSe nanosheet[27]. For the top contacted devices, a few layer of GaSe was exfoliated first on the Si/SiO$_2$ substrate, and then the metal contacts were deposited on top of it. While for the bottom contacted devices, the metal contacts were deposited first on Si/SiO$_2$ and then the GaSe was exfoliated on the metal contacts. It should be noted that the interface of the contacts between these two types of devices is slightly different, where the bottom contacted device is Ti/Au/GaSe interface while the top contacted one is GaSe/Ti (2 nm)/Au. The schematic illustrations of our devices are presented in Fig. 1a,b (see device fabrication Method). The thickness of the GaSe flakes was determined by atomic force microscopy (AFM) (see Supplementary Fig. S2 for more details). The typical thickness used for the sensitive photodetectors in this work is about ~20-30 nm. The normalized photocurrent spectrum of the photodetectors with wavelength range from 390 to 800 nm shows two well-defined peaks (Fig. 1c), where one peak is located at 412 nm corresponding to an energy gap of 3.01 eV and the other one is located at 610 nm corresponding to 2.03 eV. These two energy gaps correspond to the transition from $p_x$ and $p_y$-like orbits to the conduction band and $p_z$-like orbit to the conduction band, respectively[21]. The band gap generally increases as the thickness of layers approaches atomic dimensions for the layered materials. The monolayer has a degenerated direct and indirect band gap at 2.1 eV (590 nm) and the bulk form GaSe has a band gap at around 2.0 eV (619 nm)[19,21]. To achieve an ideal photoresponse, the wavelength at 410 nm was chosen for the following studies presented in this work.

**Photoresponsivity with global illumination.** To investigate the size and the contacting type effect on the enhancement of the photoresponsivity, photocurrent measurements based on both type devices with different spacing distances between the source and drain electrodes were carried out. Fig. 2a shows the SEM image of a device with top contacted devices and distance between each electrode from bottom to up are 0.09, 0.2, 0.5, 1, 5 and 10 μm, respectively. Current-voltage curves of devices were recorded with sweeping bias voltage under global illumination ($\lambda$ = 410 nm) with light intensity ranging from $1\times10^{-2}$ to 1.45 mWcm$^{-2}$. The photocurrent is the difference between the current under illumination and the dark current, namely $I_{ph} = I_{light} - I_{dark}$. In order to directly compare the photoresponsivity for both the bottom and top contacted devices, only the device area between the source and drain electrodes was counted for the calculation of the photoresponsivity, which is described as $R = I_{ph} / P_{light}$ with $P_{light} = WlL_{intensity}$ for the very thin nanosheet, where $L_{intensity}$ is the light intensity, W is the width of the device, and $l$ the distance between the source and drain electrodes. With the bias voltage above 2 V, the photoresponsivity is rigidly associated with the lateral spacing distance for both the bottom and top contacted devices with fixed contact width (Fig. 2b; Supplementary

Fig. S3b). With reducing the distance between the electrodes, the photoresponsivity rapidly increases at fixed $V_{DS}$ = 8 V and light intensity 0.01 mWcm$^{-2}$. However, the distance $l$ can't be cut too short since the direct tunneling will happen between the source and drain under applied bias at very small $l$, which will enlarge the dark current and reduce the sensitivity of the photodetector. To ensure the low dark current, we found that the distance between the source and drain of the photodetectors should be at least 200 nm, where the dark current starts to increase with the applied voltages above 10 V for the device with $l$ = 200 nm (Fig. 2c). Furthermore, very large dark current was observed with $V_{DS}$ above 0.2 V for the device with $l$ = 90 nm (Fig. 2c, inset). Moreover, the photoresponsivity of the bottom contacted devices increases from 200 to 5,000 AW$^{-1}$ with $l$ shrinking from 8 μm to 290 nm, while it only increases from 40 to 900 AW$^{-1}$ for the top contacted devices. Thus we can conclude that the highest photoresponsivity can be achieved in the bottom contacted photodetectors with optimized nanoscale spacing distance.

**Photoresponsivity with localized illumination.** In order to understand the nature of the photoresponsivity enhancement with shrinking the photodetector size, the localized laser beam with spot diameter of 1.5 μm was used to investigate the spatially resolved photocurrent in a rather wide top contacted device with $l$ = 9 μm (Fig. 2b,d, inset). Seven points were marked out as Point A~G with spacing about 1.5 μm between the adjacent points (Fig. 2d, inset). Independent of the bias direction, very small currents during sweeping the voltages were observed when the localized irradiating laser was located in the middle of the device, namely the marked Points C,

D, and E. However, when the laser was focused on the positions A and B, the photocurrents were obviously increased up to ten times of the current that from C~E positions under a forward bias voltage, i.e., a positive voltage is applied to the electrode near the A and B spots. Similarly, the magnitude of the photocurrents under a negative bias voltage with illumination at positions F and G were as high as those from Point A and B (Fig. 2d, inset). Measurements of photocurrents with light irradiation at different localized positions clearly demonstrated that the photocurrent is mainly generated from the photoexcited carriers close to the metal semiconductor contacts. Furthermore, the photocurrent active region is always close to the Schottky barrier with higher electrical potential.

The underlying physics of the current-voltage results with spatially localized laser illumination can be understood according to the band diagram analysis. With no illumination and drain bias voltage, the device is in its equilibrium state, characterized by Schottky barriers at the contacts. Considering GaSe as a p-type material with Fermi energy of around 5.6 eV, which is larger than the Au work function[28], we plotted the schematic band diagram of the devices (Fig. 3a,b). Illuminating the device under zero bias, with photons energy higher than bandgap, electron-hole pairs will be generated and separated in the depletion region of GaSe. However, both the photoexcited electrons and holes moved to the opposite directions at the two end Schottky barriers, which will cancel each other, as indicated in Fig. 3a. While the electron-hole pairs outside the depletion region of Schottky barrier will diffuse randomly due to the absence of electric field. As a result, the photocurrent was hardly detected under global illumination with zero bias.

Under spatially resolved illumination, the mechanism of the asymmetric photocurrent results can be divided into three situations where the localized laser was focused on the left, middle and right part of the GaSe nanosheet (taking situation with forward bias voltage as an example). For irradiation on the middle points, which is located outside of the Schottky barrier, the photogenerated electrons and holes are separated by the electric field. The electrons are drifted to the right and the holes are drifted to the left. Carriers need to travel to the metal contacts before being collected, which will mostly be recombined due to the relative small mobility (Supplementary Fig. S1) and result in a weak photocurrent. With illuminating on the right side, the built-in electric field and the electric field built by bias voltage have the same direction in the right Schottky depletion region, which will separate the photogenerated carriers more efficiently. Since holes need to drift from right side to left side, most of the photoexcited holes are scattered or recombined. Thus the photocurrent is mainly originated from the photoexcited electrons tunneling though the barriers. While with illuminating on the left side at forward bias, the built-in field in Schottky depletion region and the electric field built by the bias voltage were just in the opposite direction, which will cancel each other, thus the photogenerated carriers were separated more difficultly and tiny photocurrent was observed. Conversely the spatially resolved photocurrent shows opposite phenomenon under reverse bias, which is because of the opposite electrical potential direction and thus the band diagram.

Therefore, the asymmetric photocurrent is found to be more sensitive to the photoexcited carriers close to the Schottky barrier at the higher electrical potential side, which should also be true for the global illumination. The Schottky barrier width determines the effective absorption area. The barrier width and height together define the carrier tunneling probability. Also, the built-in electric field within the barrier and the additional electric field determine the speed of separated carriers together. The width of the Schottky barrier gets thinner under the bias voltage when the built-in electric field has the same direction to the electric field direction built by the bias voltage. These can explain why the current increases with increasing the bias voltage in the meantime at fixed light intensity.

**Discussion**

We developed a model to demonstrate the concept related to the transport of photogenerated carriers in a metal-semiconductor-metal (MSM) photodetector. By solving the continuity equations for carriers in the region of the device based on the measured device structure (see Supplementary Fig. S4 and S5 for more details), this model accurately depicts the dependence of the photoresponsivity on scale, as shown in Fig. 2b. For clarity and simplicity, at forward bias, the electrons were considered as the main carrier to generate the photocurrent. The photogenerated electrons diffuse to the interface between the GaSe and the metal contact with higher electrical potential (using $X = l$ at forward bias for example), and then the electrons have the same possibility to pass through the interface and enter into the metal contact. This model is not suitable for the extreme small devices with existing the direct tunneling between the source and drain. Luckily, the direct tunneling should be avoided in photodetectors.

For the top contacted device, the possibility of the photogenerated electrons at any arbitrary position $X = x$ reaching to the interface $X = l$ can be written down as: $\exp(-(l-x)/v\tau)$, where $v$ is the electron velocity and $\tau$ is the lifetime of the electrons. Then the total number of photogenerated electrons reached to $X = l$ per second under global illumination is

$$N = \int_0^l (\sigma W)\exp(-(l-x)/L_D)\mathrm{d}x = \sigma W L_D[1-\exp(-l/L_D)] \qquad (1)$$

where $\sigma$ is the number of the photogenerated electrons per square meter, and $L_D = v\tau$ is the diffusion length. And the photocurrent is proportional to the number of total carriers per second received at $X = l$, namely $I_{ph} = cN = c\sigma W L_D[1-\exp(-l/L_D)]$, where $c$ is a constant as the coefficient of proportionality. Under this model, the photoresponsivity thus can be written down as $R = I_{ph}/P_{light} = C_0[1-\exp(-l/L_D)]/l$, where $C_0 = c\sigma L_D/L_{intensity}$. Using this model, the spacing distance between the source drain electrodes dependence of the photoresponsivity for the top contacted devices can be well fitted using the photoresponsivity equation and the diffusion length of the electrons $L_D = 170$ nm was obtained.

However, for the bottom contacted devices, except for the photocurrent contribution described above, the photoexcited electrons in the both contacted regions also contribute to the photocurrent under global illumination. The photogenerated electrons in the left contact region have to diffuse to the right side and then enter into the metal contacts at forward bias, which can be described similarly using the above formula. However, the photoexcited electrons in right contact side will have vertical rather than planar transport and then enter into the metal contact below. Thus the photocurrent for the bottom contacted device is the sum of the planar and vertical contribution, which can be written down as:

$$I_{ph} = c \int_{-L_l}^{l} (\sigma W) \exp[-(l-x)/L_D] dx + c' \int_{0}^{d} (\sigma W L_r) \exp[-(d-x)/L_D'] dx \qquad (2)$$

where $L_l$ is the width of left contact, $c'$ is the probability of vertical transport electrons entering into the metal contact, $L_r$ is the width of the right contact, and $L_D'$ is the vertical diffusion length. Thus the photoresponsivity then can be written down as:

$$R = I_{ph}/P_{light} = C_0\{1-\exp[-(l+L_l)/L_D]\}/l + C_1[1-\exp(-d/L_D')]/l \qquad (3)$$

where the coefficient $C_1 = c'\sigma L_r L_D'/dL_{intensity}$. Normalized the experimental contacts width, the distance dependence of the photoresponsivity for the bottom contacted devices can be well described (Fig. 2b).

Taking the advantage of the bottom contacted device, we then pick out one of the typical bottom contacted photodetector with spacing $l$ = 1 μm between the two electrodes as example to carefully investigate the bias voltages, time, and photointensity dependence of the photocurrent. The optical image of the device is shown in the inset of Fig. 4a. The current as a function of the bias voltage under dark and global illumination at different irradiation intensities was shown in Fig. 4a. Very low dark current was observed in measured voltage regime, which is benefit from the device structure with two back to back Schottky barriers. Under global illumination, the current starts to increase with the applied voltages at ±2 V, which increases further with increasing the magnitude of the voltages. Also the current increases with increasing the light intensities. Current was significantly increased by two orders of magnitude, from 40 pA (dark condition) to ~6 nA at fixed light intensity ~1.7 mWcm$^{-2}$ and bias voltage 5 V. We then probed the time-dependent photorepsonse to the global illumination with light intensity 1.7 mWcm$^{-2}$ at different bias voltages (Fig. 4b). With $V_{DS}$ = 1 V, nothing was clearly observed with switching light on and off. With $V_{DS}$ above 2 V, the current sharply increases with switching on the light and drops dramatically after the light switched off, which is consistent with the current-voltage results under illumination (Fig. 4a). The sensitive, fast and reversible switching between the on and off states allows the device to act as high quality photo detectors and switchers. The dynamic response to the light illumination for rise and fall in our devices can be expressed by $I(t) = I_0[1 - \exp(t/\tau_r)]$ and $I(t) = I_0 \exp(-t/\tau_d)$, where $\tau_r$ and $\tau_d$ are the time constants for the rise and decay (Fig. 4c). The rising and falling time can be obtained by fitting the experimental results, which is shown in Fig. 4c. The photocurrent rose dramatically in 10 ms after the light illumination and decayed within 20 ms after the light-off. This is in sharp contrast to the long tails up to a few seconds after the sharp rising and falling in the previous reported few layer GaSe photodetectors, a much shorter rising and falling tails about 0.2 s with light shining on and off were observed, which is originated from cutting away the photogenerated electrons far away from the interface at the high potential side. The evaluated rising and falling speed of our photodetectors is one of the fastest among the reported data for layered material-based photodetectors[11,29].

However, this speed is still much slower than that usually observed MSM photodetectors[30], which can be attributed to the influences of traps and other defect states during the photocurrent generation processes. The fast rising and falling time and device optimization will be performed to improve the photocurrent dynamics of GaSe nanosheet devices. Important enhancements could be realized using encapsulation and surface trap state passivation.

Based on measurements of Fig. 4a,b, the light intensity dependence of the photocurrent was plotted in Fig. 4d. This can be fitted to a power law $I_{ph} \propto P^{\gamma}$, where $\gamma = 0.54$ determines the response of the photocurrent to the light intensity[31,32]. The non-unity exponent suggests a complex process of electron-hole generation, recombination and trapping within the semiconductor[31,33]. With decreasing the light intensity at fixed bias voltage $V_{DS}$ = 5 V, the corresponding photoresponsivity firstly increases and reaches the maximum of 1,200 AW$^{-1}$ (Fig. 4d, inset), which is more than 400 times higher than the previously reported GaSe photodetector[19] and five orders higher than that of graphene-based photodetectors[4,34,35]. Then the photoresponsivity decreases with increasing the photointensity for the light intensity above 0.01 mWcm$^{-2}$ (Fig. 4d, inset). This is because that the light absorption efficiency reaches to the maximum in this few-layer photodetector at relatively low photointensity of 0.01 mWcm$^{-2}$. The light intensity dependence of the photocurrent and photoresponsivity at different bias voltages was also investigated (Supplementary Fig. S3).

In summary, significant improvements in photosensitivity can be realized with shrinking the spacing distance in the layered GaSe based MSM photodetectors[19], which is more than 3 orders improvement with shrinking *l* down to 290 nm for the bottom contacted MSM photodetectors. From a broad perspective, we have developed a model for understanding the underlying physics of the photocurrent in our MSM photodetectors, which could also be widely used in any low dimensional materials based MSM photodetectors. Our work suggests that it is feasible to design bottom contacted nanoscale MSM photodetectors based on layered materials with very high photoresponsivity, which will open pathways for future integrated optoelectronic applications.

**Methods**

**Device fabrication.** The GaSe nanosheet photodetectors used in this work were prepared by mechanical exfoliation of CVD growth GaSe single crystal. GaSe flakes were identified by optical microscope and their thicknesses were further confirmed by AFM. Devices with two types of contacts were fabricated: bottom contacted electrodes and top contacted electrodes. We firstly pre-patterned the alignment marks using optical lithography on a $SiO_2$(300 nm)/$Si^{++}$ substrate. For the top contacted device, firstly the few layer of GaSe was exfoliated on the $Si/SiO_2$ substrate, and then the metal contacts Ti/Au (2/80 nm) were deposited using thermal evaporator. While for the bottom contacted devices, the metal contacts Ti/Au (2/40 nm) with designed width 700 nm were thermally evaporated first and then the GaSe was exfoliated on it. SEM images in the manuscript were performed using a JEOL JSM6510 operated at 20 KV with $LaP_6$ filament.

**Electrical measurements.** All electrical and optoelectrical measurements were measured using Agilent Technology B1500A under vacuum of $10^{-6}$ mbar at room temperature. The as-prepared samples behaved in a p-type manner from FET results. GaSe nanosheet sample have a very low mobility as $5\times10^{-3}$ cm$^2$ V$^{-1}$s$^{-1}$.

**Global illumination measurements.** Monochromatic illumination was provided by a Zolix Omni-λ300 monochrometer with a Fianium WhiteLase Supercontinuum Laser Source with repetition rate 20 MHz. The output laser wavelength can be tuned by monochromator Omni-λ 300. The laser beams could directly irradiate the nanosheet device through a transparent glass window of the vacuum chamber. The laser spot size is about 1 mm$^2$ on the sample for the optoelectrical measurements under global illumination.

**Spatially resolved photocurrent measurements.** A microscope objective and a micromechanical stage were used to localize the corresponding position of the focused laser beam on the photodetector, where the diameter of the laser spot size was about 1.5 μm and the illumination power was fixed at 1 μW. The current-voltage (*I-V*) measurements were performed with the spatially resolved localized laser from A to G positions.

**Acknowledgements**

This work was supported by "973 Program" Nos. 2011CB922200 and 2014CB643900, NSFC Grant No. 61225021. K. W. also acknowledges the support of Chinese Academy of Sciences "100 talent program".

**Author contributions**

K. W. conceived and designed the experiments. P. Hu provided the high quality CVD growth GaSe single crystals. Y. C. fabricated all the devices. T. Y. helped for the spatially resolved measurements. Y. C. and K. C. performed the measurements. Y. C., K. C., X. W. and K. W. co-wrote the paper. All authors discussed and commented on the manuscript.

**Additional Information**

The authors declare no competing financial interests. The structure, electrical and the details of the model were shown in the supplementary information. Correspondence should be addressed to K. W. (kywang@semi.ac.cn).

# Figure captions

**Figure 1 | Photodetector structure.** **(a)** A schematic of the photodetector with the contacts at the top. **(b)** A schematic of the photodetector with the contacts at the bottom. **(c)** The normalized photocurrent of the GaSe photodetector as a function of the illumination wavelength.

**Figure 2 | Both top and bottom contacted photodetectors with different spacing distance.** **(a)** The scanning electron microscopy image of the typical top contacted MSM photodetectors with Scale bar of 5 μm. The smallest spacing distances between the metal fingers is 90 nm and the finger width is 700 nm. **(b)** The photoresponsivity as a function of the spacing distances at $V_{DS}$=8 V for both the top contacted (red) and bottom contacted (blue) photodetectors, where the dash lines are the fitting results using our models. The direct tunneling is appeared under bias in the grey area with $l \leq 200$ μm, which will decrease the photoresponsivity. **(c)** Dark current voltage characteristics for the photodetectors with different spacing distances. **(d)** Current voltage characteristic of spatially resolved localized illumination. The up left inset shows the device image and the position of illumination. Bottom right inset shows the spots of the illumination.

**Figure 3 | Schematic band diagrams of the MSM devices**. **(a)** Band diagram of the photodetector with zero bias voltage under global illumination. **(b)** Band diagram of the photodetector with forward bias voltage under global illumination.

**Figure 4 | Bottom contacted photodetector with 1 μm spacing distance**. **(a)** Photocurrent as a function of the drain voltage under global illumination with different light intensities at fixed wavelength of 410 nm. Inset shows the optical image of the device. **(b)** Time-resolved photoresponse of the photodetector, recorded for different bias voltages $V_{DS}$ with fixed light intensity $P_{light}$ = 1.7 mWcm$^{-2}$ . The period of the laser on and off is 20 seconds. **(c)** The rise and decay of the normalized photocurrent at the initial stage just after the laser is switched on (upper panel) and off (lower panel), where the dots are the experimental results and the dash dots are the fitting results. **(d)** Photocurrent as a function of the light intensity at fixed bias voltage $V_{DS}$ = 5 V, where the red line is the fitting result. Inset shows the light intensity dependence of the photoresponsivity at fixed bias voltage $V_{DS}$ = 5 V.

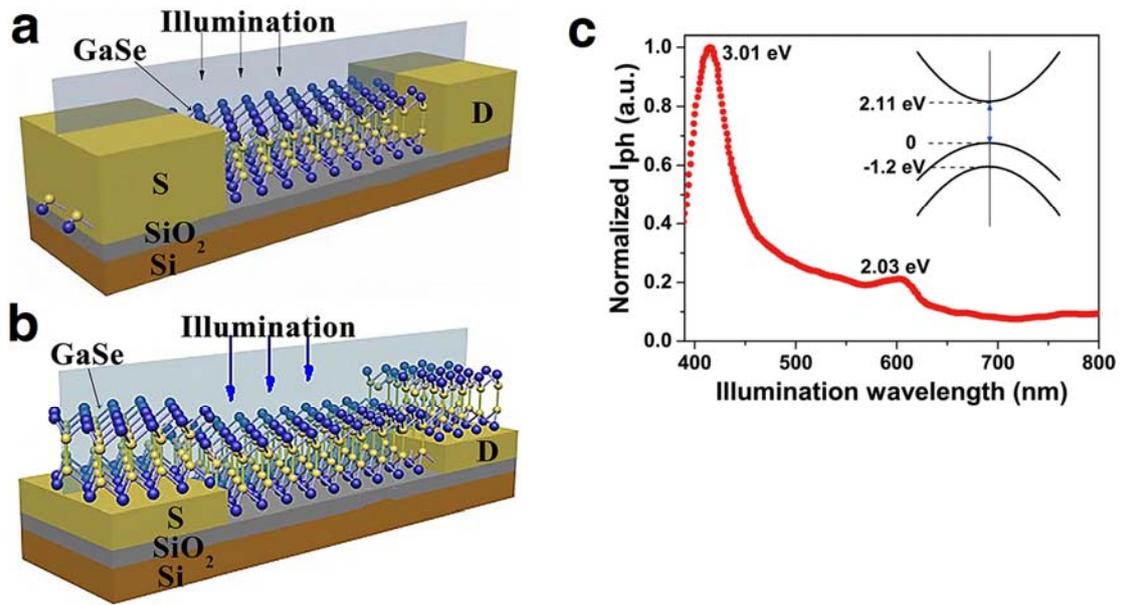

*Figure 1 Y. F. Cao et al.*

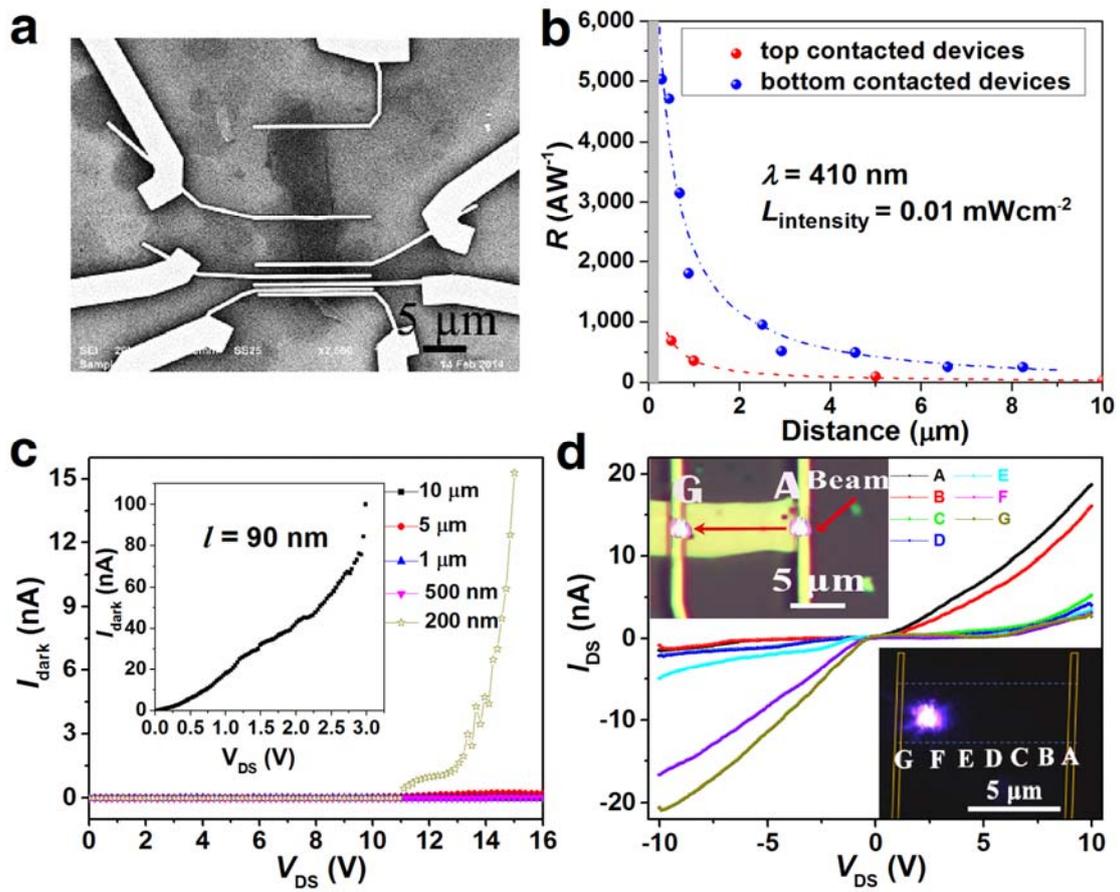

*Figure 2 Y. F. Cao et al.*

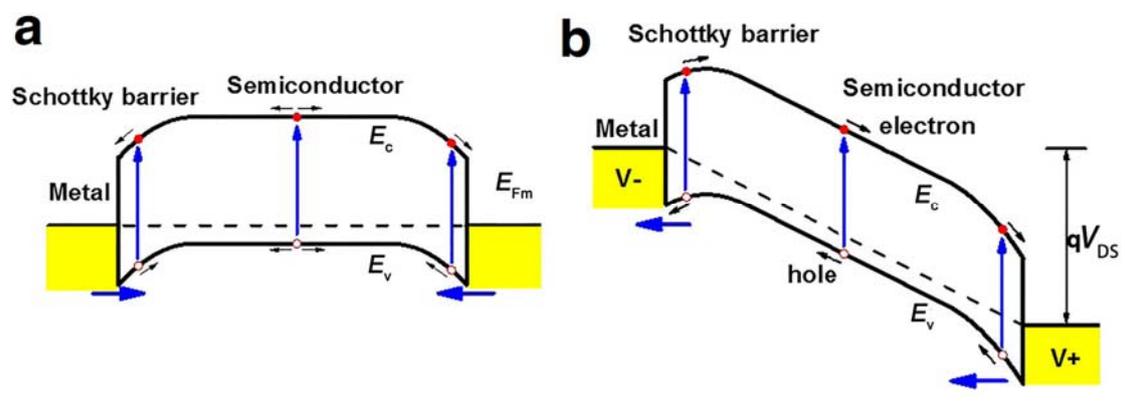

*Figure 3 Y. F. Cao et al.*

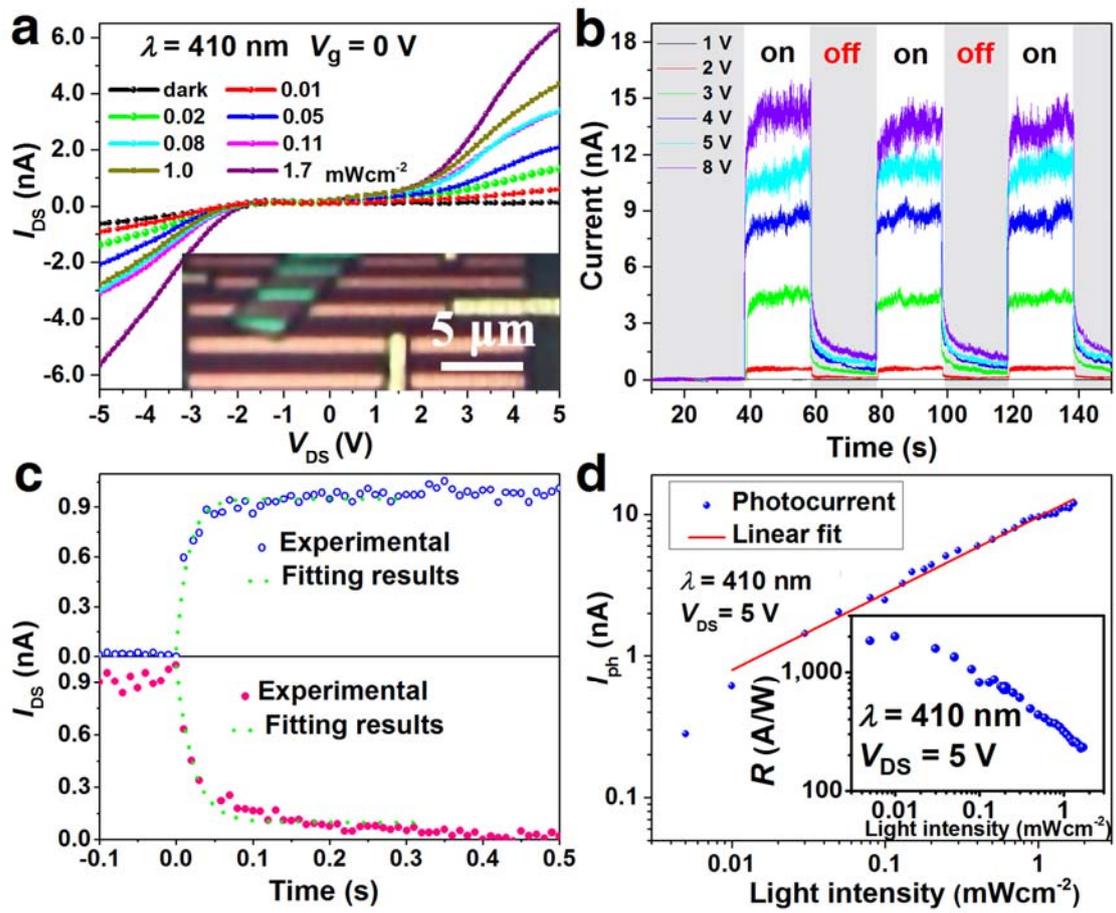

*Figure 4 Y. F. Cao et al.*